\documentclass[aip,apl,twocolumn,superscriptaddress]{revtex4}

\usepackage{graphicx}
\usepackage[ansinew]{inputenc}
\usepackage{array}
\usepackage{color}
\usepackage{amsmath}
\usepackage{amsxtra}
\usepackage{amstext}
\usepackage{amssymb}
\usepackage{latexsym}
\begin{document}

\title{Response of graphene to femtosecond high-intensity laser irradiation}

\author{Adam Roberts}
\email{aroberts@optics.arizona.edu}
\affiliation{College of Optical Sciences, University of Arizona, Tucson, AZ, 85721 USA.}

\author{Daniel Cormode}
\author{Collin Reynolds}
\author{Ty Newhouse-Illige}
\author{Brian J. LeRoy}
\affiliation{Department of Physics, University of Arizona, Tucson, AZ, 85721 USA.}

\author{Arvinder Sandhu}
\email{sandhu@physics.arizona.edu}
\affiliation{Department of Physics, University of Arizona, Tucson, AZ, 85721 USA.}
\affiliation{College of Optical Sciences, University of Arizona, Tucson, AZ, 85721 USA.}

\date{\today}

\begin{abstract}
We study the response of graphene to high-intensity 10$^{10}$-10$^{12}$ Wcm$^{-2}$, 50-femtosecond laser pulse excitation. We establish that graphene has a fairly high (I$_{th} \sim 3\times10^{12}$ Wcm$^{-2}$) single-shot damage threshold. Above this threshold, a single laser pulse cleanly ablates graphene, leaving microscopically defined edges. Below this threshold, we observe laser-induced defect formation that leads to degradation of the lattice over multiple exposures. We identify the lattice modification processes through in-situ Raman microscopy. The effective lifetime of CVD graphene under femtosecond near-IR irradiation and its dependence on laser intensity is determined. These results also define the limits of non-linear applications of graphene in femtosecond high-intensity regime.
\end{abstract}

\maketitle
Graphene is a two-dimensional form of carbon that exhibits novel physical properties due to its unique lattice and electronic band-structure\cite{Novoselov2005}. While many recent investigations have focused on the optical and electronic applications of graphene\cite{Bonaccorso2010}, attention has also been devoted to graphene's non-linear properties. Important demonstrations of graphene's non-linear response include the generation of mode-locked ultrafast laser pulses \cite{Sun2010}, non-linear four-wave mixing and harmonic generation\cite{Hendry2010, Dean2009}.  In such non-linear applications, the graphene lattice interacts with intense, ultrashort light pulses. The light absorption and dissipation mechanisms in this regime form an active topic of investigation\cite{Wang2010,Sun2008}. One aspect of this line of inquiry relates to the thresholds and limits for graphene's non-linear optical response. Clearly, there exists an upper limit for the photon flux that this unique single-atom thick carbon layer can withstand. Prior work has shown that the graphene lattice can be significantly modified by high doses of continuous-wave (CW) laser irradiation and electron beam irradiation\cite{Krauss2009,Teweldebrhan2009}. However, unlike CW irradiation, a strong impulsive excitation by an ultrashort pulse produces a very different response in graphene that needs to be understood further.

The graphene used in our measurements was grown on copper foil using the CVD process and then transferred to thin glass substrates\cite{Li2009}. The ultrafast pulses were generated in a Ti:Sa laser amplifier with a center wavelength of 790nm and focused to  20-30 micron spot on graphene. We controlled the pulse energy, pulse duration and the number of pulses used to irradiate the graphene sample. Importantly, we utilize in-situ Raman spectroscopy to probe the laser exposed samples with $\mu$m  resolution. Raman measurements were performed with a 532nm laser at low power, so that the probing does not modify the graphene lattice. 

\begin{figure}[t]
\includegraphics[width=0.49 \textwidth]{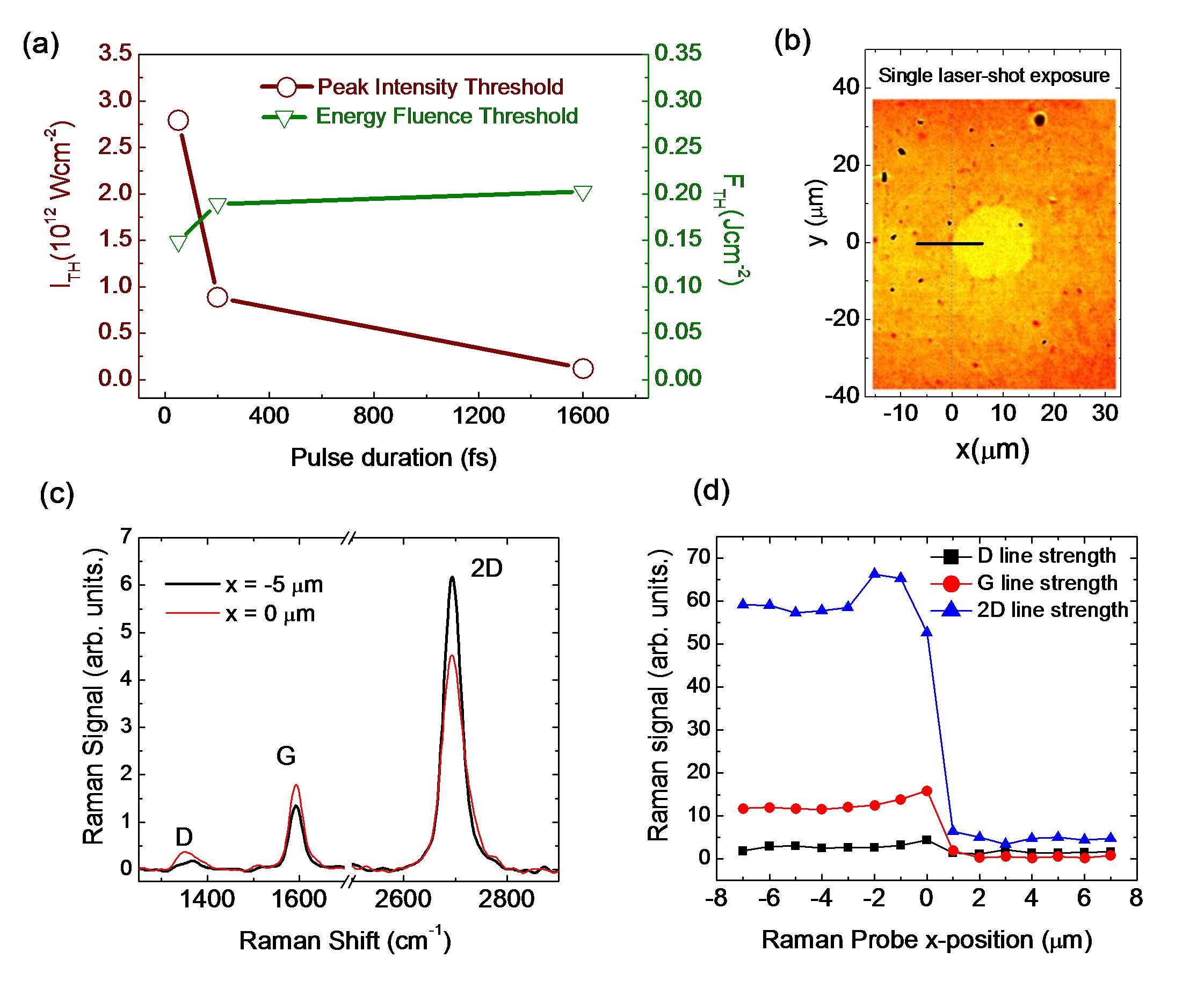}
\caption{\label{fig:1} (a) Intensity and Fluence thresholds for single-shot laser damage of graphene. (b) Optical image of femtosecond pulse damage spot. (c) Raman spectra at two different locations, near and away from the edge of laser damage spot. (d) Raman probe scan across the edge (black line in (b)) of damage spot. Raman line strengths are defined as area under the spectral peaks. }
\end{figure}

As a first step we obtain a single-shot laser damage threshold of graphene. We define the damage threshold as the point at which a single laser pulse exposure creates a hole in the carbon lattice. The damage threshold was obtained for different pulse durations as shown in  Figure \ref{fig:1}(a).  In the range from 50 fs to 1.6 ps, the energy fluence at which the graphene damaged was nearly the same$F_{TH} \sim 200$mJ/cm$^2$.  This compares well with the theoretically predicted threshold of  250mJ/cm$^2$\cite{Jeschke2001} for ultrafast damage of a graphitic film.  In terms of peak intensities, graphene survives under significantly higher intensity for a shorter pulse. At 50fs duration the intensity threshold is $I_{TH} \sim 2.7 \times 10^{12}$ Wcm$^{-2}$. Notably, the femtosecond damage threshold is much higher in comparison to the point at which CW laser leads to lattice modification, which is observed to be $\sim10^6$ Wcm$^{-2}$ \cite{Krauss2009}.

We observe that single-shot damage threshold for femtosecond pulses is very well-defined. The lattice survives without much modification up to a certain intensity value, but beyond that value it is completely ablated.  Figure \ref{fig:1}(b) shows an optical transmission image of a graphene layer exposed to a single 50fs laser pulse above the damage threshold. The contrast between the graphene-covered and graphene-free areas is evident. 

To characterize the microscopic lattice modification we use Raman probing. Figure \ref{fig:1}(c) shows the Raman spectra obtained at two positions along $y=0$ line of the optical image in Fig.  \ref{fig:1}(b). Raman spectra consist of D, G, and 2D lines, where the D line strength (i.e. area under the spectral peak) is a measure of defects in the lattice \cite{Malard2009}.  Outside of ablated area (x=-5$\mu$m), we observe a predominantly mono-layer lattice \cite{Graf2007} with very few defects. Near the boundary (x=0$\mu$m), we observe increase in D peak due to edge defects. 

Figure 1(d) shows the Raman probe scan along the solid black line (at y=0) superimposed on the optical image in Fig. \ref{fig:1}(b). We observe a very sharp transition for D, G, and 2D Raman line strengths at x=0$\mu$m boundary between the ablated and the graphene-covered area. Outside of the ablated area ($x<0$), where the lattice gets exposed to laser intensities below the damage threshold, graphene survives well and stays close to being pristine.  However, inside of the ablated area the Raman peaks fall to the level of the background noise. The sharpness of the transition region size in our measurements is only limited by the Raman probe spot-size of 2$\mu$m. Observation of a sharp transition between the unmodified and ablated region offers the interesting possibility of using femtosecond laser pulses to micro-machine graphene patterns. 

\begin{figure}[t]
\includegraphics[width=0.49 \textwidth]{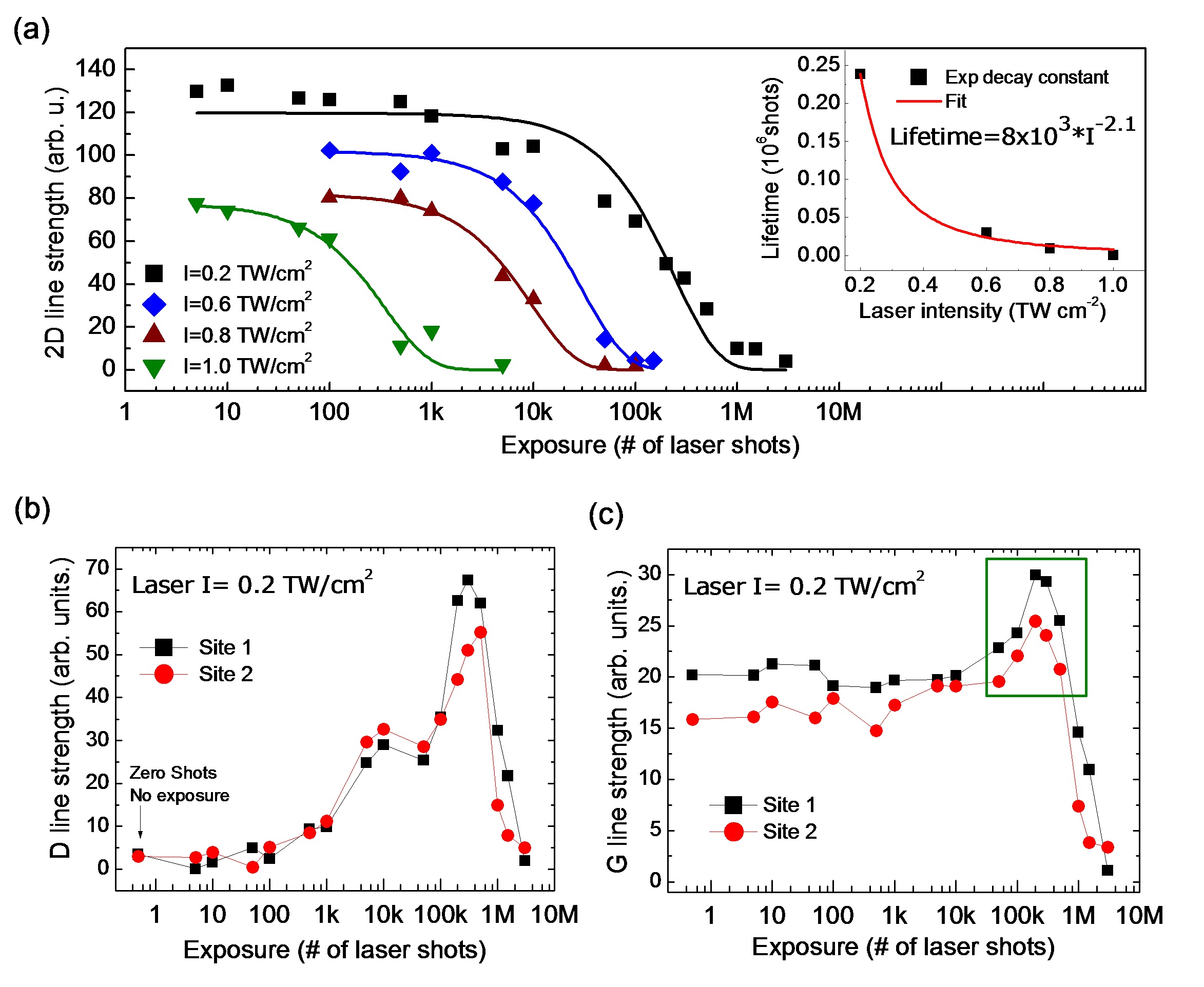}
\caption{\label{fig:2}(a) Evolution of 2D line strength as a function of number of laser shots and exponential decay fits at various peak intensities. Inset: Dependence of lifetime on peak laser intensity. (b) Variation of the disorder related D-line with multiple laser exposures at 2x10$^{11}$Wcm$^{-2}$. (c) Variation of G-line strength with laser exposure. Raman signal shows increase in the high disorder region, before decaying to zero. }
\end{figure}

Next, we measured the modification of the graphene lattice due to multiple exposures below the single-shot damage threshold.  These studies were conducted for the 50fs laser pulses at four different peak intensities in the range between $0.2-1 \times 10^{12}$ Wcm$^{-2}$. Lattice modification was deduced from variation of Raman spectra as a function of the number of laser exposures. Figure\ref{fig:2}(a) plots the evolution of Raman 2D-line strength (i.e. area under the 2D peak) with laser exposures. The variation of 2D signal was fit with a decaying exponential  $S_{2D} \sim \exp[-N/N_o(I)]$, where $S_{2D}$ is the strength of 2D line,$N$ is the number of exposures, and$N_o(I)$ is the decay constant which depends on the intensity of the ultrafast pulses.  From $S_{2D}$ fits we obtain the decay lifetimes (in terms of laser exposures) which depends on the laser intensity as $N_o(I) = \alpha \times 8 \times 10^3 I^{-2.1}$ where $I$ is the peak laser intensity in the units of $TWcm^{-2}$ (Inset of Fig \ref{fig:2}(a)).  Extrapolating from this empirical relationship between lifetime and laser intensity, our sample should have a decay lifetime of about 10$^{8}$ exposures for a laser intensity of 10$^{10}$ Wcm$^{-2}$. At 1kHz repetition rate, this corresponds to 1.5 days of continuous laser exposure, before the graphene lattice shows degradation. Our preliminary studies also indicate that these lifetimes are longer when using exfoliated graphene instead of CVD graphene.

While the 2D signal can characterize the sample lifetime for particular pulse intensity, the onset of disorder in the lattice is directly seen by monitoring the D line as shown in figure \ref{fig:2}(b).  With increasing laser exposures, the Raman D-line increases in strength until it reaches a saturation limit and then falls. As discussed below, this behavior is related to the number of defects that are accumulating in the lattice\cite{Ferrari2007} and the inter-defect distance. Another interesting feature of laser modification of graphene is observed through the variation of G-phonon line strength that is shown figure  \ref{fig:2}(c). Near the point at which the D line strength has peaked, the G line strength shows an unexpected increase.  Clearly, this increase cannot be attributed to an increase in the number of scattering centers (i.e. carbon-carbon bonds), so it must be due to a process that increases the Raman scattering cross section. In the discussion below, we explain the interesting mechanisms at play. It should also be noted that our data is very repeatable over different sites on the CVD graphene sample. 

\begin{figure}[t]
\includegraphics[width=0.49 \textwidth]{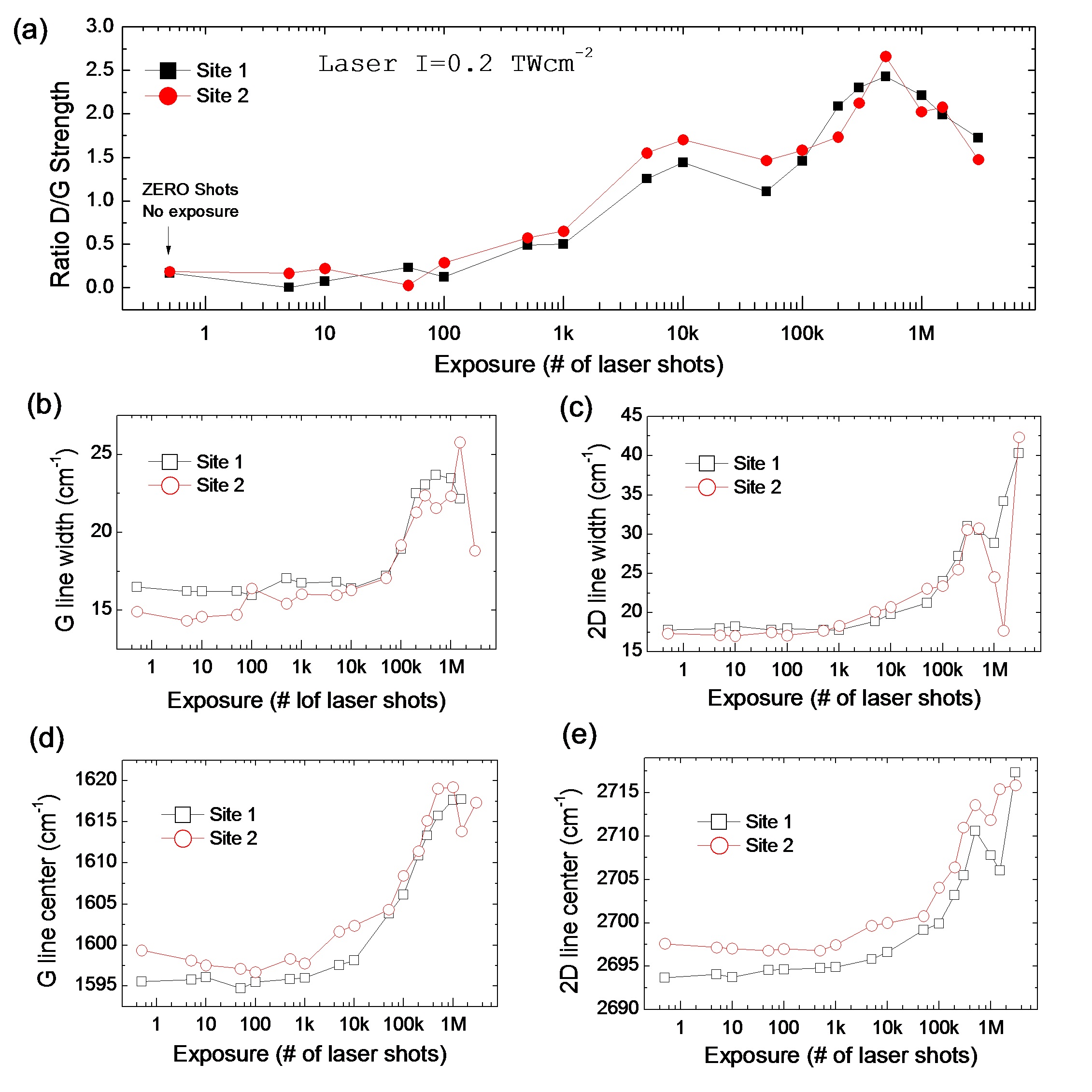}
\caption{\label{fig:3} (a) Ratio D/G line strength as a function of number of laser exposures at 2x10$^{11}$ Wcm$^{-2}$. (b) and (c) show the increase of G and 2D line widths, respectively. (d) and (e) plot the shift of G and 2D Raman lines with laser exposures.}
\end{figure}

Figure \ref{fig:3}(a) examines the ratio of D to G line strength, which can be used to quantify the disorder in the lattice. Our ratio increases with laser exposures until it reaches about 2.3, and then falls. This can be understood based on discussion of disorder in graphene by Ferrari\cite{Ferrari2007}. With the progression of laser irradiation, the ultrafast excitation breaks carbon bonds, thus increasing the number of defects in the lattice which in turn leads to the formation of smaller and smaller nano-crystallites. In this regime, the ratio of D to G line strength ($S_D/S_G$ ) is inversely proportional to an average inter-defect distance or nano-crystallite size $L_{a}$, which can be express as $S_D/S_G = C_1/L_a$, where $C_1$ is a constant. However, there is a saturation limit for the defects, where most of the sample is in the form of amorphous sp$^2$ carbon. At this point, breaking any more bonds will open up sp$^2$ carbon rings, decreasing the sites available for the double resonance process that is associated with D phonon scattering, thus reducing the D-line Raman signal.  This regime is modeled as $S_D/S_G = C_2L_a^2$. $C_1 \sim 4.4$ nm and $C_2 \sim 0.0055$ nm$^{-2}$ from the literature\cite{Ferrari2007,Tuinstra1970},  the $S_D/S_G$ peak ratio of 2.3 corresponds to an average graphene nano-crystallite size $L_a \sim 2$ nm, which agrees well with the observations reported in \cite{Ferrari2007}.  These observations are also consistent with disorder induced increase in Raman widths illustrated in Figs.  \ref{fig:3}(b) and \ref{fig:3}(c).

Now we turn our attention to the unexpected G-peak increase shown in the boxed region of figure \ref{fig:2}(c)). Recently, Chen et al. \cite{Chen2011} reported enhancement of Raman G-line signal due to heavy p-doping of graphene. Essentially, the net Raman signal arises from quantum interferences between several G-phonon-scattering pathways. Heavy doping can move the Fermi energy to the point that some the destructively interfering pathways are cut off, leading to the overall enhancement of signal. For this to occur, the Fermi energy had to be close to half of the Raman excitation photon energy. We believe that this effect comes into play in our experiment due to adsorption of atmospheric dopants to the dangling carbon bonds associated with laser induced defect sites. Since our G peak enhancement occurs when the nano-crystallite size approaches 2nm, the number of defects where dopants can attach is very high. The resulting heavy doping is very likely to produce large negative Fermi energy shifts which in turn lead to Raman G line enhancement. 

Assuming a linear relationship between Fermi energy, $E_F$, and frequency shift, $\Delta \omega_G$, as observed in  \cite{Chen2011}, $\Delta \omega_G = |E_F| \times 42$ cm$^{-1}$ eV$^{-1}$, the 25 cm$^{-1}$ line shift corresponds to a Fermi energy shift of 0.6 eV. This shift is seen to produce a 50$\%$ Raman enhancement in G line strength in ref.\cite{Chen2011}, which is similar to what we observe in our case. Lastly, we can estimate the carrier (dopant) density corresponding to this Fermi energy shift. Using $n = (E_F/\hbar v_F)^2/\pi$\cite{Novoselov2005}, we obtain carrier concentrations of $n \sim3 \times10^{13}$ cm$^{-2}$.  Further, we note that the 2D line is also blueshifted (Fig. \ref{fig:3}(e)) which indicates that dopants in our case are p-type\cite{Das2008,Yan2007}. This large p-type doping could occur due to atmospheric oxygen attaching to the dangling bonds on the fragmented lattice\cite{Ryu2011}.

In conclusion, we  identified a distinct single-shot damage threshold ($\sim3TWcm^{-2}$) for CVD grown monolayer graphene when exposed to an intense 50 fs laser pulse. The edges from single-shot laser ablation were found to be microscopically clean, which indicates potential for ultrafast laser micro-patterning of graphene.   Below the single-shot damage limit, ultrafast pulse exposures lead to the formation of defects, which transforms pristine graphene into nano-crystallites whose size can be determined from the ratio of D to G line strength. The defect sites accumulate p-type dopants, which manifests in the form of blueshifts of the G and 2D lines. The decay of the 2D Raman peak upon laser irradiation was used to obtain the relationship between lifetime and peak laser intensity. Our results indicate that for CVD graphene, a safe working regime for femtosecond pump-probe studies and non-linear applications is $<10^{10}$Wcm$^{-2}$.

We thank the TRIF photonics and DoD SMART program for providing graduate scholarship funding. We also acknowledge support from the NSF REU program, NSF EECS/0925152 and the U. S. Army Research Office under contract W911NF-09-1-0333.



\end{document}